\documentclass{aa}  

\usepackage{graphicx}
\usepackage[normalem]{ulem}
\usepackage[dvipsnames]{color}
\usepackage{txfonts}
\usepackage{array}
\usepackage{float}
\usepackage{natbib}
\newcommand{\Msol}{$\mathrm{M_{\odot}}$}

\newcommand{\kms}{$\mathrm{km} \, \mathrm{s}^{-1}$}

\newcommand{\elc}{L17\textit{c}}
\newcommand{\magsqarcsec}{$\mathrm{mag}\,\mathrm{arcsec}^{-2}$}

\newcommand{\mhi}{$M_\mathrm{HI}$}
\newcommand{\wfifty}{$W_{50}$}

\begin{document}

\title{The contribution of HI-bearing ultra-diffuse galaxies to the cosmic number density of galaxies}

\author{
M.~G. Jones\inst{\ref{iaa}}, 
E. Papastergis\inst{\ref{kapteyn},\ref{rabo}},
V. Pandya\inst{\ref{ucsc}},
L. Leisman\inst{\ref{cornell},\ref{valpo}}, 
A.~J. Romanowsky\inst{\ref{sjsu},\ref{uco}},
L.~Y.~A. Yung\inst{\ref{rutgers}},
R.~S. Somerville\inst{\ref{rutgers},\ref{flat}},
\and
E.~A.~K. Adams\inst{\ref{astron},\ref{kapteyn}}
}

\institute{
Instituto de Astrof\'{i}sica de Andaluc\'{i}a, Glorieta de la Astronom\'{i}a, Granada 18008, Spain \label{iaa}
\and
Kapteyn Astronomical Institute, University of Groningen, Landleven 12, Groningen NL-9747AD, The Netherlands \label{kapteyn}
\and
Credit Risk Modeling Department, Co\"{o}perative Rabobank U.A., Croeselaan 18, Utrecht NL-3521CB, The Netherlands \label{rabo} 
\and
Department of Astronomy \& Astrophysics, University of California Santa Cruz, 1156 High Street, Santa Cruz, CA 95064, USA \label{ucsc}
\and
Cornell Center for Astrophysics and Planetary Science, Space Sciences Building, Cornell University, Ithaca, NY 14853, USA \label{cornell}
\and
Department of Physics \& Astronomy, Valparaiso University, Valparaiso, IN 46383, USA \label{valpo}
\and
Department of Physics \& Astronomy, San Jos\'{e} State University, One Washington Square, San Jose, CA 95192, USA  \label{sjsu}
\and
University of California Observatories, 1156 High Street, Santa Cruz, CA 95064, USA \label{uco}
\and
Department of Physics and Astronomy, Rutgers University, 136 Frelinghuysen Road, Piscataway, NJ 08854, USA \label{rutgers}
\and
Center for Computational Astrophysics, Flatiron Institute, 162 5th Ave, New York, NY 10010, USA
\label{flat}
\and
ASTRON, the Netherlands Institute for Radio Astronomy, Postbus 2, Dwingeloo NL-7900AA, The Netherlands \label{astron}
}

\titlerunning{The number density of HI-bearing UDGs}

\authorrunning{M.~G. Jones et al.}

\abstract{
We estimate the cosmic number density of the recently identified class of HI-bearing ultra-diffuse sources (HUDs) based on the completeness limits of the ALFALFA survey. These objects have HI masses approximately in the range $8.5 < \log M_{\mathrm{HI}}/\mathrm{M_{\odot}} < 9.5$, average $r$-band surface brightnesses fainter than 24 \magsqarcsec, half-light radii greater than 1.5 kpc, and are separated from neighbours by at least 350 kpc. In this work we demonstrate that they contribute at most $\sim$6\% of the population of HI-bearing dwarfs detected by ALFALFA (with similar HI masses), have a total cosmic number density of $(1.5 \pm 0.6) \times 10^{-3}$ $\mathrm{Mpc^{-3}}$, and an HI mass density of $(6.0 \pm 0.8) \times 10^{5}$ $\mathrm{M_{\odot}\,Mpc^{-3}}$. We estimate that this is similar to the total cosmic number density of ultra-diffuse galaxies (UDGs) in groups and clusters, and conclude that the relation between the number of UDGs hosted in a halo and the halo mass, must have a break below $M_{200} \sim 10^{12}$ \Msol \ in order to account for the abundance of HUDs in the field. The distribution of the velocity widths of HUDs rises steeply towards low values, indicating a preference for slow rotation rates compared to the global HI-rich dwarf population. These objects were already included in previous measurements of the HI mass function, but have been absent from measurements of the galaxy stellar mass function owing to their low surface brightness. However, we estimate that due to their low number density, their inclusion would constitute a correction of less than 1\%. Comparison with the Santa Cruz semi-analytic model shows that it produces HI-rich central UDGs that have similar colours to HUDs, but these are currently produced in much great a number. While previous results from this sample have favoured formation scenarios where HUDs form in high spin parameter halos, comparisons with the results of \citet{Rong+2017}, which invokes that formation mechanism, reveal that this model produces an order of magnitude more field UDGs than we observe in the HUD population, and these have an occurrence rate (relative to other dwarfs) that is approximately double what we observe. In addition, the colours of HUDs are bluer than those predicted by \citet{Rong+2017}, although we suspect this is due to a systematic problem in reproducing the star formation histories of low-mass galaxies rather than being specific to the ultra-diffuse nature of these sources.
}

   %\keywords{ --  -- }

\maketitle

\section{Introduction}
\label{sec:intro}

\defcitealias{Leisman+2017}{L17}

Over the past few years there has been enormous interest in ultra-diffuse galaxies \citep[UDGs,][]{vDokkum+2015a}, a population of highly extended, yet low-mass, galaxies that have been identified predominantly in groups and clusters. These objects have radii typical of $L^{\ast}$ galaxies, but stellar masses of dwarf galaxies. While low surface brightness galaxies have been studied for decades \citep[e.g.][]{Sandage+1984,Schombert+1992}, including some that are now classified as UDGs, the prevalence of this population in clusters, even at the lowest surface brightnesses, was not previously recognised. Their formation mechanism and how they can survive as undisturbed objects in the cluster environments, where most have been found \citep[e.g.][]{vDokkum+2015a,Koda+2015,Mihos+2015,vdBurg+2016,Yagi+2016}, remain unsolved problems, although several formation scenarios have now been proposed \citep[e.g.][]{Amorisco+Loeb2016,Rong+2017,DICintio+2017}.

A smaller number of galaxies with similar optical properties to the cluster UDGs have been identified in lower density environments \citep[e.g.][]{Martinez-Delgado+2016,Roman+2017,Bellazzini+2017,Trujillo+2017,Makarov+2015}. \citet[][hereafter L17]{Leisman+2017} also identified a field population of 115 HI-bearing ultra-diffuse galaxies (HUDs) with the HI survey, ALFALFA \citep[Arecibo L-band Fast ALFA,][]{Giovanelli+2005,Haynes+2011}, that were selected to have equivalent properties to the \citet{vdBurg+2016} UDGs. It is unclear whether this gas-rich UDG population is directly related to the cluster UDG population, as might be expected under some formation mechanisms, or if they are a separate population with a different formation mechanism or mechanisms \citep[see also][]{Papastergis+2017}.

An essential step in establishing the nature of the HUD and UDG populations is to measure their cosmic abundance, which can then be used to inform and constrain potential formation models. \citet{vdBurg+2017} estimated the abundances of these objects in clusters and groups, and \citet{Karachentsev+2017} concluded that only 1.5\% of galaxies in the Local Volume ($D < 11$ Mpc) are potential UDGs, but in the field such an accounting has not yet been possible. Although a number of UDGs have been detected in the field, the lack of both a blind method of detection over a wide area \citep[although such efforts are now in progress, e.g.][]{Greco+2017} and a means to measure redshifts has presented a challenge. In this work we take the HUDs sample of \citetalias{Leisman+2017} and estimate their number density within the Local Universe, based on the completeness limits of the ALFALFA survey. This is the first measurement of its kind for UDGs in the field, and we use it to assess what contribution these objects make to the HI mass function (HIMF) and to galaxy stellar mass function (SMF), and, where possible, to make comparisons with the predictions of existing formation scenarios.

This paper is arranged as follows: Section \ref{sec:datasets} describes the sample of \citetalias{Leisman+2017}, Section \ref{sec:method} explains the method used to estimate the cosmic abundance of HUDs, and Section \ref{sec:SAM} describes the Santa Cruz SAM (semi-analytic model) with which we compare. The results are presented in Section \ref{sec:results} and discussed in Section \ref{sec:discussion}. We draw our conclusions in Section \ref{sec:conclusions}. Throughout this paper we used the value of the Hubble constant as $H_{0} = 70 \; \mathrm{km\,s^{-1}\,Mpc^{-1}}$, the matter density of the Universe as $\Omega_{\mathrm{m,0}} = 0.27$, and the critical density as $\rho_{\mathrm{c,0}} = 2.75 \times 10^{11} \; \mathrm{M_{\odot}\,Mpc^{-3}}$.

\section{HI sample}
\label{sec:datasets}

We use the \citetalias{Leisman+2017} sample of gas-bearing and isolated UDGs. The \citetalias{Leisman+2017} HUDs were identified within the dataset of the ALFALFA blind HI survey \citep[][]{Giovanelli+2005,Haynes+2011}. In particular, \citetalias{Leisman+2017} searched for HI sources within the 70\% ALFALFA catalogue\footnote{The 70\% ALFALFA catalogue can be accessed at \url{http://egg.astro.cornell.edu/alfalfa/data/index.php}.}, which covers 70\% of the final ALFALFA footprint, that have optical properties that are consistent with various definitions of UDGs in the literature. Full details of the selection process can be found in \citetalias{Leisman+2017}, but here we summarize the main selection criteria:

\begin{enumerate}

\item \textit{High-quality ALFALFA detections.} Sources are selected based on a high signal-to-noise ratio, $S/N_{\mathrm{HI}} \geq 6.5$, and confident classification as extragalactic objects (``code 1'' in the ALFALFA catalogue). \label{item:quality}

\item \textit{Available SDSS imaging data.} Sources must be located within the imaging footprint of the Sloan Digital Sky Survey (SDSS) DR12 . In addition, HI sources that are located within $10^\prime$ of a bright star in the Yale Bright Star Catalog are excluded, because their optical photometry could be compromised. \label{item:counterpart}

\item \textit{Distance limits.} Sources are required to have distances 25 Mpc $< D <$ 120 Mpc.\footnote{Distances to sources were taken from the ALFALFA catalogue, which were calculated from a local Universe flow model \citep{Masters2005} combined with assignments to the Virgo cluster and groups \citep[see][and Jones et al. (in preparation) for more details]{Hallenbeck+2012}.} The upper distance limit is imposed to mitigate source blending, as the physical size corresponding to the $\sim 3.5^\prime$ ALFALFA beam grows linearly with distance. The lower distance limit is placed to reduce uncertainties in source distances, as peculiar motions have a larger fractional contribution to the recessional velocity for nearby objects.   \label{item:distance}

\item \textit{Isolation criteria.} Sources are selected to have no neighbours in the Arecibo General Catalog (AGC) that fall within 350 kpc projected separation on the plane of the sky and $\pm 500$ \kms \ in recessional velocity. The AGC is a database which includes all known redshifts in the Local Universe as available in NED, including all optical redshifts from SDSS and HI redshifts from ALFALFA. It is maintained by M.~P. Haynes and R. Giovanelli. \label{item:isolation}

\item \textit{Low optical surface brightnesses and large half-light radii.} The selection of HI sources with UDG-like optical properties is implemented in two steps: First, objects with relatively high surface brightness are excluded based on SDSS pipeline photometry in the $g$, $r$, and $i$ bands. More specifically, HI sources whose SDSS counterpart has an average surface brightness within the exponential radius, \texttt{expRad}, brighter than 23.8 \magsqarcsec \ (in any of the three bands), or an average surface brightness within the radius including 90\% of the Petrosian flux, \texttt{petroR90}, brighter than 25 \magsqarcsec, are excluded. The resulting 645 candidate sources are then visually inspected to remove objects with erroneous SDSS pipeline photometry. 

Finally, for the remaining $\sim$200 candidates, \citetalias{Leisman+2017} performed manual photometry on the SDSS images. These measurements are used to select a ``restrictive'' sample of 30 sources with optical properties equivalent to those of the \citet{vDokkum+2015a} UDGs, and a ``broad'' sample of 115 sources equivalent to the \citet{vdBurg+2016} definition of UDGs. The ``restrictive'' sample contains the lowest surface brightness objects, but the ``broad'' sample is still fainter than the majority of ``classical'' low surface brightness objects (e.g. Figure 1 of \citetalias{Leisman+2017}). The ALFALFA sources (of similar HI masses and velocity widths to HUDs) that are excluded by these criteria appear to be either fairly typical dwarf galaxies (nearly face-on) or, in some cases, low surface brightness objects that are excluded due to brighter regions such as star-formation clumps (this is discussed further in Section \ref{sec:caveats}).

As calculation of the HIMF requires many objects across several mass bins, we consider only the ``broad'' sample in this work in order to minimise the Poisson uncertainties. Thus, the HUDs discussed in this paper have the following requirements on their optical properties: average surface brightness $\langle \mu_{r}(<r_{\mathrm{eff},r}) \rangle > 24$ \magsqarcsec, half-light radius $r_{\mathrm{eff},r} > 1.5$ kpc, and absolute magnitude $M_{r} > -17.6$ mag. \label{item:low_sb}

\end{enumerate}

In addition to the HUD sample of \citetalias{Leisman+2017}, we will also make use of the overall population of HI sources from the 70\% catalogue of the ALFALFA survey (hereafter $\alpha$.70). The $\alpha$.70 catalogue contains a total of 18987 high signal-to-noise ($S/N > 6.5$) extragalactic HI sources, and the completeness and reliability of the ALFALFA survey have been thoroughly quantified in Section 6 of \citet{Haynes+2011}. All ALFALFA sources are ultimately extracted by a person leading to almost 100\% reliability for high signal-to-noise sources, and in this paper we make use of the ALFALFA completeness limit, which we approximate as a step function at the position of the 50\% limit \citep[Equations 4 \& 5 from][]{Haynes+2011}.

\section{Analysis method}
\label{sec:method}

\begin{table}
\centering
\caption{Sample counts after successive cuts}
\label{tab:counts}
\begin{tabular}{lcc}
\hline\hline
Cuts                 & HUDs & $\alpha$.70  \\ \hline
Distance \& Isolation            & 115  & 5225 \\
$S/N > 6.5$ & 115  & 4500 \\
Completeness limit   & 102  & 4318 \\ \hline
\end{tabular}
\tablefoot{The number of sources in the HUD sample and $\alpha$.70 with successive cuts applied. Note that the first two cuts are already part of the definition of a HUD so do not remove any sources from that sample. The full $\alpha$.70 HIMF is calculated from the 16620 sources above the completeness and signal-to-noise limits that are available without the distance and isolation cuts (or the requirement to fall within the SDSS footprint).}
\end{table}

Due to the complicated selection criteria of the \citetalias{Leisman+2017} sample of HUDs, a straightforward computation of their cosmic number density is not possible. For example, the isolation criteria (item \ref{item:isolation} in the preceding section) make it very difficult to specify the volume over which HUDs of different HI masses have been detected and, as a result, the calculation of the absolute normalization of their number density is very challenging. In addition, the \citetalias{Leisman+2017} HUDs are preferentially located in relatively low-density environments because of the isolation criteria and, as a result, the spatial distribution of \citetalias{Leisman+2017} HUDs deviates substantially from the underlying large scale structure in the survey volume. Consequently, computing accurate volume correction factors that take into account large-scale density fluctuations in the survey volume is not possible. For this reason, we follow an indirect approach in calculating the cosmic number density of the gas-bearing UDGs, and their contribution to the overall number density of galaxies.

First we exclude from both the \citetalias{Leisman+2017} HUD sample and the $\alpha$.70 catalogue those sources whose observed HI flux falls below the 50\% completeness limit of the ALFALFA survey (consult Equations 4 \& 5 in \citealt{Haynes+2011}). Out of the 115 HUDs in \citetalias{Leisman+2017} sample, 13 are rejected due to this criterion, leaving 102. We refer to this 102-member sample as the ``\elc'' sample (\textit{c} stands for cut).

Second, we apply to all sources in the $\alpha$.70 catalogue the same distance and isolation criteria used to define the \citetalias{Leisman+2017} sample of HUDs (see Table \ref{tab:counts}). This means that only $\alpha$.70 sources with 25 Mpc < D < 120 Mpc are considered, and that $\alpha$.70 sources that have a neighbour within 350 kpc projected distance and $\pm$500 \kms \ in recessional velocity, or those within $10^\prime$ of a bright star, are excluded (refer to items \ref{item:distance} \& \ref{item:isolation} in the preceding section). In this way, we create a sample of ``normal'' ALFALFA sources that share the exact same selection criteria as the \elc \ sample of UDGs. This sample consists of 4318 objects and is hereafter referred to as the ``$\alpha$.70$c$'' sample. By definition, this sample contains all of the \elc \ HUDs.

Next, we compute the \textit{ratio} of the number density of the \elc \ and $\alpha$.70$c$ samples. This ratio is calculated within bins of HI mass, \mhi, and within bins of HI-line profile velocity width, \wfifty \ (see Figure 1 in \citealt{Papastergis+2015} for a graphical illustration of \wfifty). These ratios refer to the fractional contributions of HUDs to the galactic HI mass function (HIMF) and velocity width function (WF), respectively. The HIMF and WF are the number density of galaxies as a function of \mhi \ and \wfifty. Last, we derive the HIMF and WF of the \citetalias{Leisman+2017} sample of HUDs by applying the ratios above to the HIMF and WF measurements obtained from the full $\alpha$.70 sample \citep{Papastergis+Shankar2016}. If $n(M_\mathrm{HI})$ denotes the HIMF, then our calculation can be summarized as

\begin{eqnarray}
n(M_\mathrm{HI})_\mathrm{HUD} = \frac{n(M_\mathrm{HI})_{\mathrm{L17}c}}{n(M_\mathrm{HI})_{\mathrm{\alpha.70}c}} \times n(M_\mathrm{HI})_\mathrm{\alpha.70} \;\;\; .
\end{eqnarray}

\noindent
The reason for following this indirect, two-step approach for calculating the number density of the \citetalias{Leisman+2017} sample is the following: First, taking the number density ratio between the two identically selected \elc \ and $\alpha$.70$c$ samples allows us to cancel in large part the biases that the selection criteria of the UDG sample induce on the measurement of its HIMF and WF. We can then use the robust HIMF and WF measurements from the $\alpha$.70 sample as references in order to obtain a much more reliable HIMF and WF for the HUD sample.

An important caveat of this approach is that by applying the same isolation criteria to the full $\alpha$.70 sample we are implicitly assuming that the isolation of HUDs in the \citetalias{Leisman+2017} sample is not a property that is intrinsic to the galaxies themselves, but was instead applied in \citetalias{Leisman+2017} purely to assist in the characterisation of the sample. In other words, we are assuming that HUDs are found in similar environments to the rest of the HI-rich population. This point will be further explored in the upcoming work of Janowiecki et al. (in preparation), but the preliminary findings suggest this assumption is valid. However, if this assumption were to be incorrect then the estimated abundances of HUDs could decrease by up to a factor of approximately 2.

All number densities in this article are calculated via the ``$1/V_\mathrm{eff}$'' method \citep{Zwaan+2003,Martin+2010}, which is a non-parametric maximum likelihood technique. Full details of the implementation of the technique for the ALFALFA dataset can be found in Appendix A of \citet{Papastergis+2015} and references therein, but here we describe it very schematically using the HIMF as an example:    

\begin{eqnarray}
n_j \equiv n(M_\mathrm{HI} = M_j) \equiv \left. \frac{\mathrm{d}N_{\mathrm{gal}}}{\mathrm{d}V \, \mathrm{d}\log_{10}(M_{\mathrm{HI}})} \right\vert_{M_{\mathrm{HI}} = M_j} \;\;\; .
\label{eqn:himf}
\end{eqnarray}

\noindent
In the equation above, $n_j$ is the value of the HIMF in the $j^\mathrm{th}$ (logarithmic) bin of HI mass centered on \mhi$= M_j$, $\mathrm{d}\log_{10}(M_{\mathrm{HI}})$ is the logarithmic width of each HI mass bin, while $\mathrm{d}V$ is a small volume element that is representative of the cosmic average, lastly, $\mathrm{d}N_{\mathrm{gal}}$ is the number of galaxies found within the volume $\mathrm{d}V$ that have HI masses within the logarithmic bin centered on $M_j$. In practice, HIMF values are calculated by counting the number of survey detections in a given logarithmic mass bin and correcting each count by an ``effective volume'' term, $V_\mathrm{eff}$, as follows  

\begin{eqnarray}
n_j  = \frac{1}{\mathrm{d}\log_{10}(M_{\mathrm{HI}})} \; \sum^i_{\mathrm{gal.}\,i\,\in\,\mathrm{bin}\,j} \frac{1}{V_{\mathrm{eff},i}}
%, \; \mathrm{for}\, \mathrm{every}\, \mathrm{galaxy}\, i \, \mathrm{in}\, \mathrm{mass}\, \mathrm{bin}\,j. 
\;\;\; .
\label{eqn:Veff}
\end{eqnarray}

\noindent
Here the summation runs over every galaxy $i$ that belongs to mass bin $j$. The effective volume, $V_{\mathrm{eff},i}$, corresponds to the total volume over which source $i$ would be detectable by the survey, but rescaled to take into account the relative overdensity or underdensity of that volume compared to the cosmic average. One technical difficulty in the computation of Equation \ref{eqn:Veff} is that an accurate computation of the effective volumes requires a sample that traces well the underlying large-scale structure (refer to Appendix B in \citealp{Martin+2010}). The \elc \ and $\alpha$.70$c$ samples do not satisfy this requirement by definition, since they consist of isolated objects only. As a result, the \textit{full} $\alpha$.70 sample (i.e., without isolation and distance cuts applied) is used to compute $V_\mathrm{eff}$ values for the galaxies in the cut samples.

\section{Model comparison}
\label{sec:SAM}

We compare our observational results with the Santa Cruz SAM of galaxy formation, which includes prescriptions for the hierarchical growth of structure, gas heating and cooling, star formation, stellar evolution, feedback from massive stars and supernovae, chemical evolution, feedback from supermassive black holes (SMBHs), and the structural and morphological transformations of galaxies due to mergers. The Santa Cruz SAM has a long history \citep{Somerville+Primack1999,Somerville+2001} and has undergone many upgrades for galactic disc formation \citep{Somerville+2008a}, SMBH feedback \citep{Somerville+2008b}, diffuse interstellar dust \citep{Somerville+2012}, mergers and disc instabilities \citep{Porter+2014}, and multi-phase gas \citep{Somerville+2015}. Studies have shown that this SAM reproduces well many properties of observed local galaxies \citep[e.g.,][]{Lu+2014,Somerville+Dave2015} as well as galaxies observed out to moderately high redshift \citep[$z<3$; e.g.,][]{Brennan+2015,Brennan+2017,Pandya+2017}. Here we use the \citet{Somerville+2015} version of the SAM which predicts HI masses, has been re-calibrated to the Planck 2013 cosmology \citep{Planck_2014}, and includes a refined treatment of star formation (Yung et al., in preparation).  

Since we want to explore UDGs with stellar masses as low as $\sim$10$^6$ \Msol, here we run the SAM using merger trees that are constructed with the Extended Press-Schechter formalism \citep[e.g.,][]{Somerville+Kolatt1999}. While we expect only minimal differences with merger trees extracted directly from N-body dark matter (DM) simulations (at least for the particular properties we consider in this paper), we do not have direct measures of environment. As a proxy for the isolation criteria applied in the observations, we restrict our analysis only to ``central" galaxies (i.e., we discard all satellites). This is justified given that the isolated HUDs in the observations are likely almost all central galaxies, unlike the classical red Coma cluster UDGs \citep{vDokkum+2015a}.

For details regarding the modelling of various physical processes in the SAM, we refer the reader to the references above. For this paper, the most important aspect is the size modelling (as this determines if a give galaxy is ultra-diffuse), which we briefly review here \citep[see also Section 3.3 of][]{Somerville+2018}. The SAM predicts separately the stellar disc exponential scale length based on angular momentum conservation \citep{Somerville+2008b} and the spheroid size based on energy conservation and virial theorem arguments \citep[see][]{Porter+2014}. We compute the 3D composite half-mass stellar radius as the stellar mass-weighted sum of the disc and bulge radii \citep[Equation 10 of][]{Porter+2014}. We then convert that 3D composite half-mass radius to a 2D projected half-light radius separately for disc-dominated and bulge-dominated galaxies (based on their bulge-to-total stellar mass ratio) following the simple procedure described in \citet{Somerville+2018}. Note that the 2D projected half-light radii are in the rest-frame $V$-band; we assume that the half-light radius remains constant across the three optical bandpasses considered in this paper (SDSS $gri$).

The stellar size--mass relation of the SAM at $z\sim0.1$ is higher in the median by a factor of $\sim$1.5--2 compared to observations \citep{Brennan+2017,Pandya+2017}. Since this offset can result in artificially diffuse galaxies, we re-calibrate the median SAM stellar size--mass relation to match the median GAMA $z\sim0.1$ size--mass relation \citep[using Table B1 of][which does not require any division based on color or morphology]{Lange+2015}. One caveat is that the \citet{Lange+2015} GAMA size--mass relation is a single power law with a completeness limit of $M_*\sim2.5\times10^9$ \Msol \ whereas we are going down to $M_*=10^6$ \Msol. Furthermore, Figure 6 of \citet{Lange+2015} suggests that at low stellar masses, the size--mass relation of red early-type galaxies shows an upturn that would not be captured by a single unbroken power law (in contrast, their low-mass blue galaxies appear consistent with a single power law). A potential adverse impact of this correction factor might be to underestimate the number density of HUDs in the SAM, especially at very low stellar masses. However, as we will show below, we still over-predict the number of HUDs at low stellar masses even with this correction. 

With this correction applied we then identify UDGs in the SAM by enforcing the same selection criteria as in the observations: the mean surface brightness within the effective radius in the SDSS $gri$ bands must be $>24$ mag arcsec$^{-2}$, the effective radius itself must be $>1.5$ kpc, and the integrated absolute magnitude $M_r>-17.6$ (AB mag). For the SMF, we separately create an HI-rich UDG sub-sample (to mimic the requirement of HUDs to be HI-bearing) by requiring $M_{\rm HI}>10^8$ \Msol. While this may seem like an oversimplification, with the $V_{\mathrm{eff}}$ weighting method discussed in the previous section, the HUD HIMF and SMF that we will calculate from the HI observations will be representative of all field UDGs with $\log M_{\mathrm{HI}}/\mathrm{M_{\odot}} > 8.2$ (barring any additional selection effects, see Section \ref{sec:caveats}). Finally, as described above, we only consider central galaxies in the SAM to mimic the isolation criterion for the observations. Throughout, we will also plot results for our ``parent” SAM sample which includes all central galaxies regardless of their surface brightness, effective radius and HI mass.

%\vspace*{2cm}

\section{Results}
\label{sec:results}

\begin{figure}
\centering
\includegraphics[width=\columnwidth]{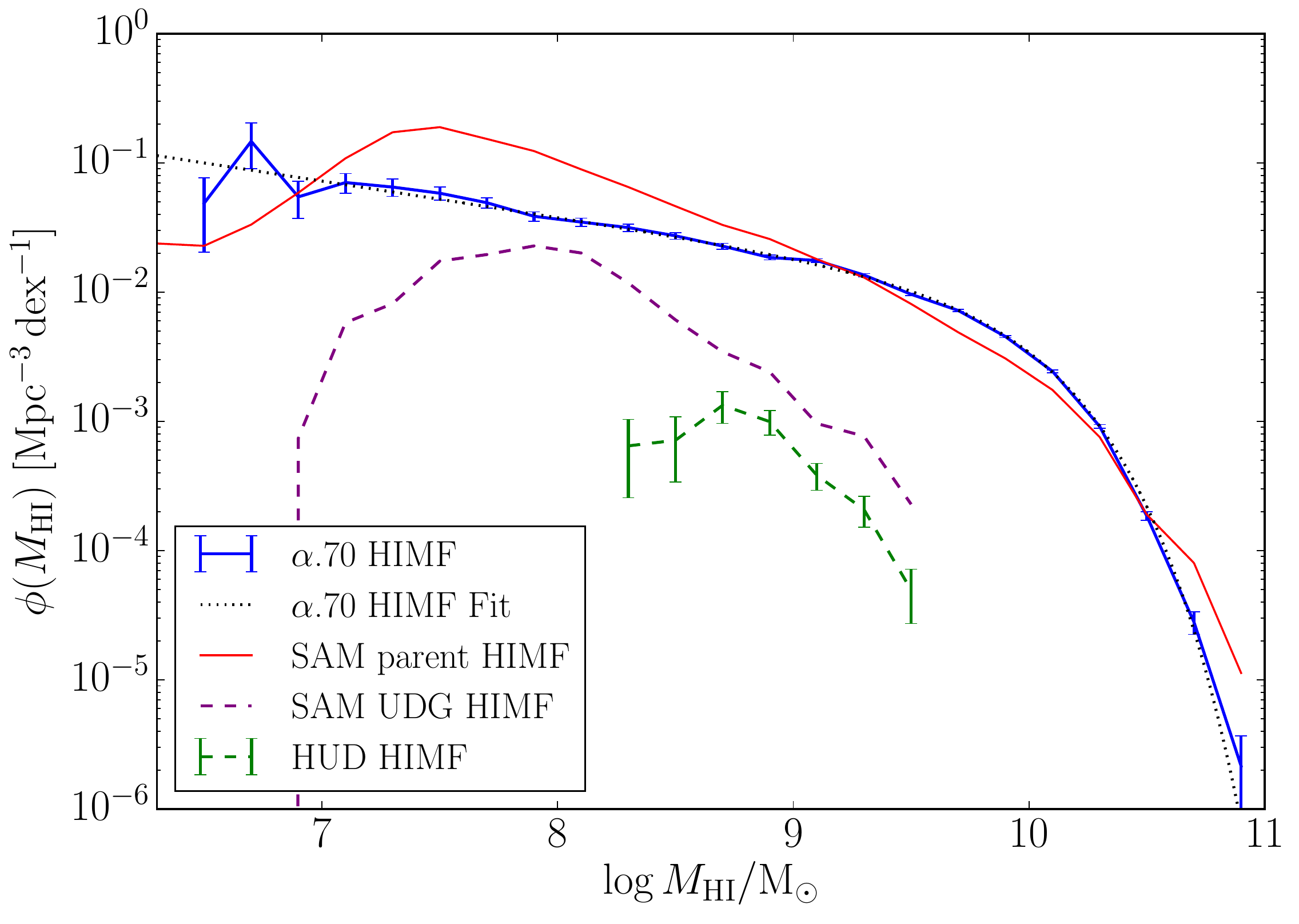}
\includegraphics[width=\columnwidth]{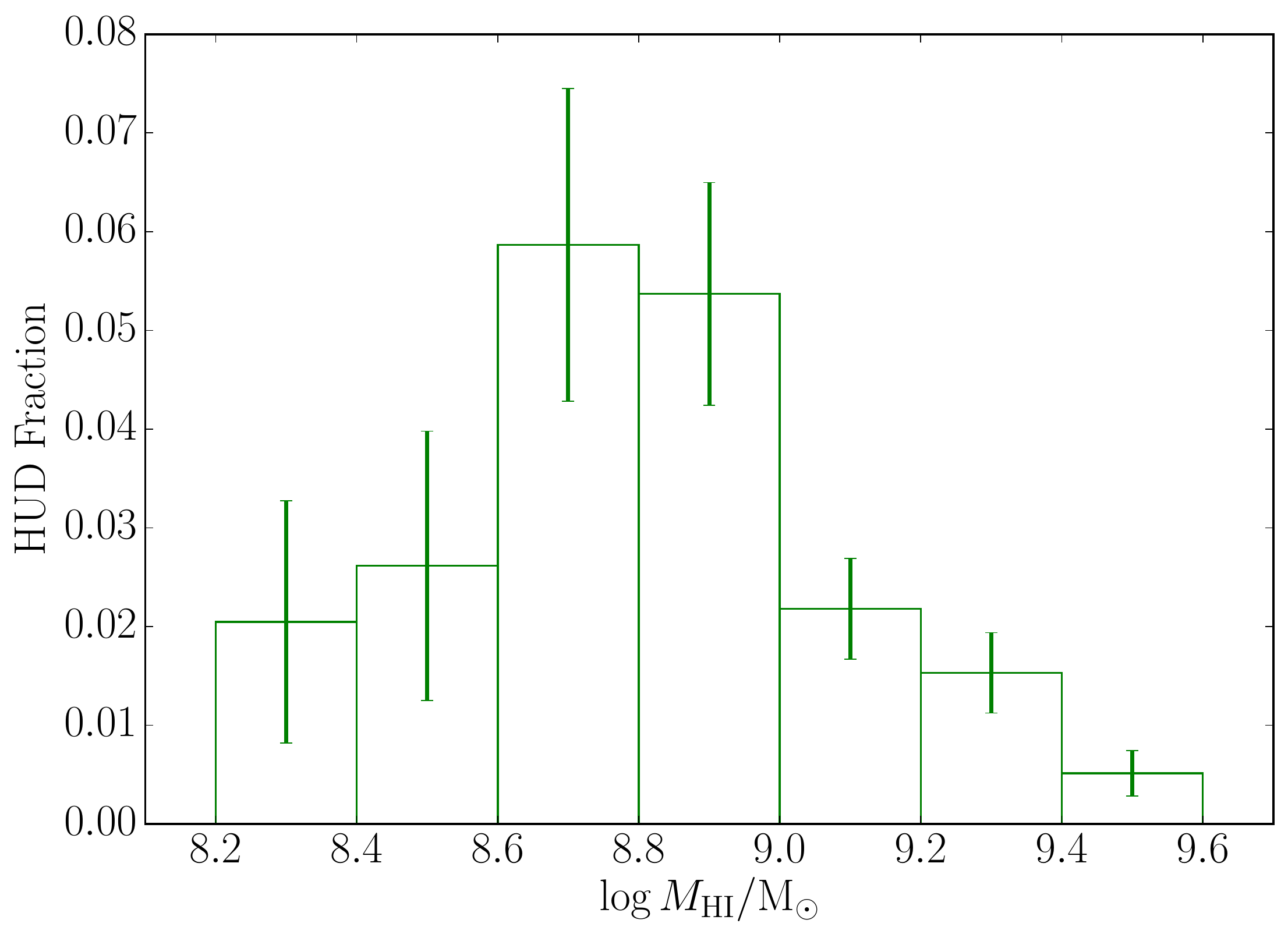}
\caption{
\textit{Top}: The HUD HIMF (dashed green line and errorbars) calculated via our ratios approach (see the text) compared to the full $\alpha$.70 HIMF (blue solid line and errorbars) and its Schechter function fit (dotted black line). We also show for comparison the Santa Cruz SAM HIMF for all galaxies (red solid line) and central UDG HIMF (dashed purple line). \textit{Bottom}: The fraction of ALFALFA HI sources that are HUDs with all appropriate weightings applied. Error bars in both plots are 1-$\sigma$ and only account for Poisson uncertainties. Bins that contain only one object have been removed.
}
\label{fig:hud_himf}
\end{figure}

\begin{figure}
\centering
\includegraphics[width=\columnwidth]{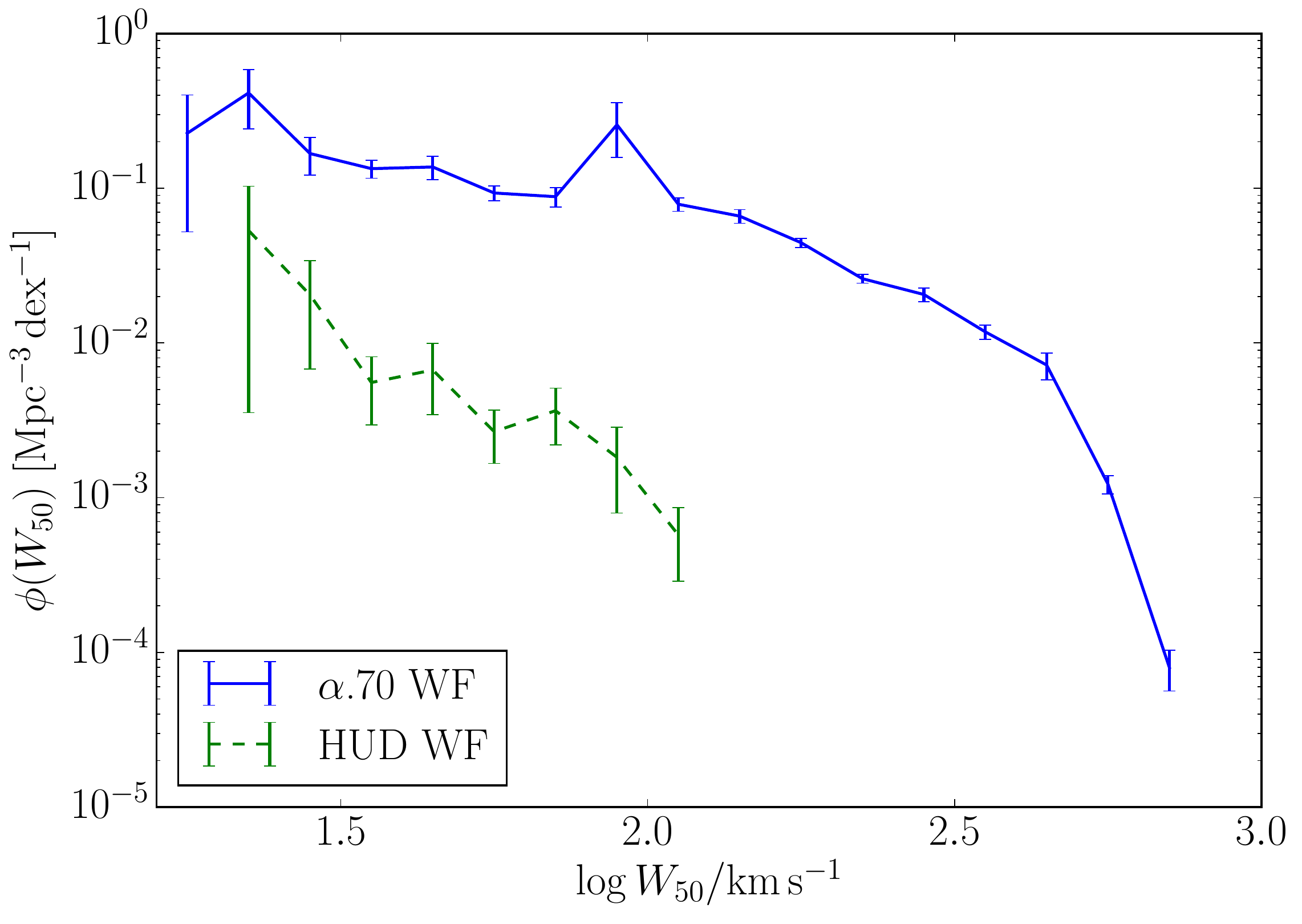}
\includegraphics[width=\columnwidth]{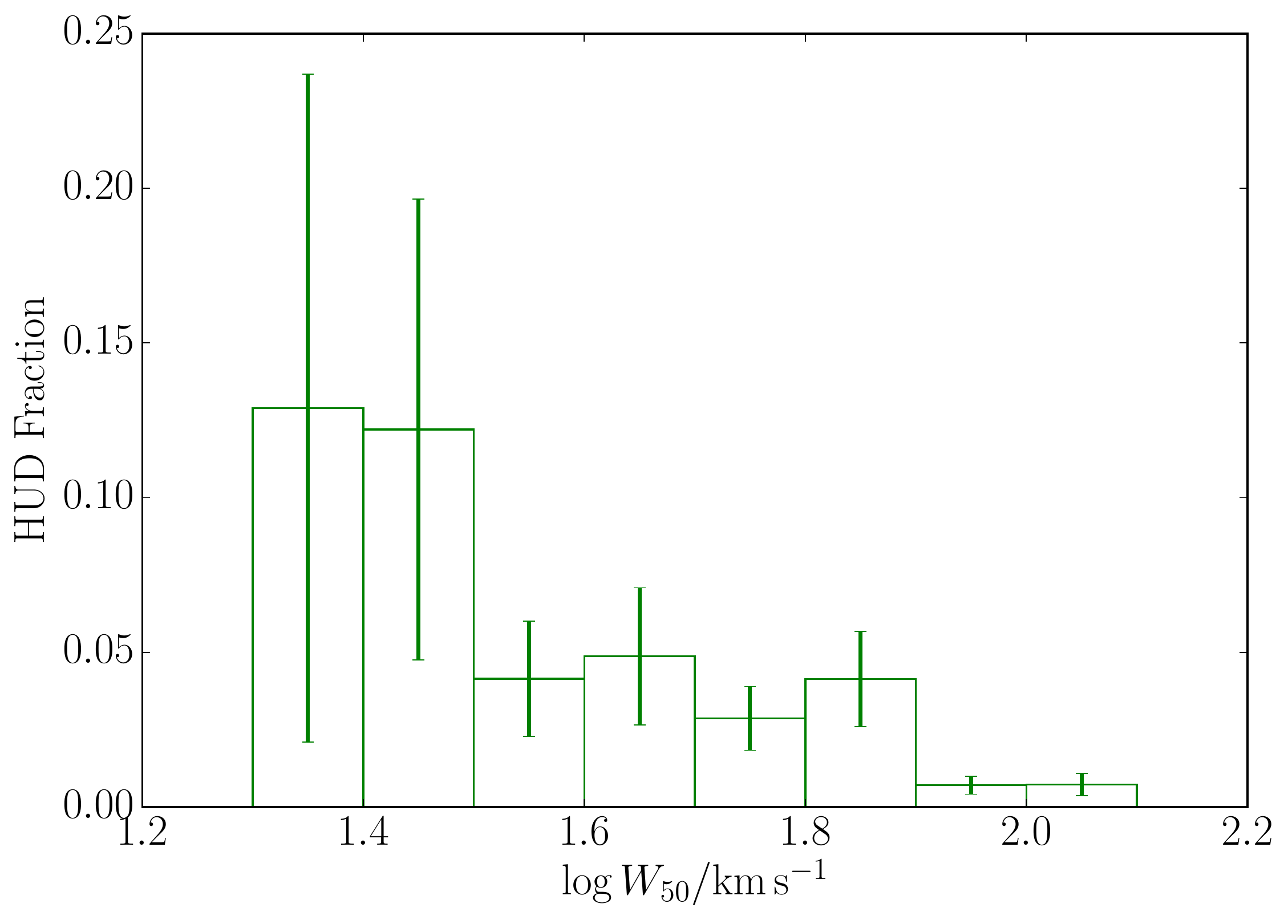}
\caption{
As for figure \ref{fig:hud_himf}, except for the HI velocity width function (and without SAM comparisons).
}
\label{fig:hud_wf}
\end{figure}

Figures \ref{fig:hud_himf} and \ref{fig:hud_wf} show the HUD galaxy HIMF and WF respectively, compared to those of the full $\alpha$.70 catalogue and to the Santa Cruz SAM. Considering first the HIMF (Figure \ref{fig:hud_himf}) it is clear that the HUD population is a minor contributor to the global HIMF at all masses. Essentially the entire HUD population has HI masses in the range $8.5 < \log M_{\mathrm{HI}}/\mathrm{M_{\odot}} < 9.5$. This may seem to be a selection effect in that $\sim$70\% of ALFALFA sources with $\log M_{\mathrm{HI}}/\mathrm{M_{\odot}} < 8.5$ are detected within 25 Mpc, and $\sim$70\% with $\log M_{\mathrm{HI}}/\mathrm{M_{\odot}} > 9.5$ are detected beyond 120 Mpc. However, as the HUD HIMF was calculated using the ratio of abundances for an equivalently selected ALFALFA sample, the fractional abundance of HUDs should not be subject to strong selection effects where there are data. This indicates that the fractional abundance of HUDs peaks at a mass of $\log M_{\mathrm{HI}}/\mathrm{M_{\odot}} \sim 8.8$, making up about 6\% of the HI galaxy population at that mass, and declines towards both higher and lower HI masses. Therefore, expanding the distance criteria of \citetalias{Leisman+2017}'s sample should not substantially alter these results. The SAM UDGs on the other hand are found to be substantially more numerous at all HI masses, and the distribution does not turn over until much lower masses. In addition, this turn over at low masses appears to be a feature of the parent population in the SAM, not something specific to UDGS.

The situation for the WF is somewhat different (Figure \ref{fig:hud_wf}). Instead of being an intermediate population, as HUDs appear to be in terms of their HI masses, HUDs are concentrated at narrow velocity widths, with the distribution continually rising towards narrower widths with a much steeper gradient than the overall HI population. The first bin ($1.2 < \log W_{50}/\mathrm{km\,s^{-1}} < 1.3$) would have appeared to continue this trend, but was removed because it only contained 1 source.

Numerically integrating the HUD HIMF gives the total cosmic number density of HUDs as $(1.5 \pm 0.6) \times 10^{-3}$ Mpc$^{-3}$, and calculating the first moment gives their HI mass density as $(6.0 \pm 0.8) \times 10^{5}$ $\mathrm{M_{\odot}\,Mpc^{-3}}$. The difference in fractional uncertainties is due to the error in number density being dominated by the lowest mass sources, which have much less impact when weighted by HI mass. HUDs therefore represent approximately 5\% of the number density of all galaxies with $\log M_{\mathrm{HI}}/\mathrm{M_{\odot}} > 8$, and less than 1\% of the total cosmic density of HI (which is dominated by the HI content of $L^{*}$ galaxies).

\section{Discussion}
\label{sec:discussion}

The results of the previous Section are the first accounting of the number density of UDGs outside of clusters. This has not previously been possible as there are not currently large area, blind, optical surveys that have the capability to reliably detect these objects in the field and measure their redshifts. However, our analysis bears the strong caveat that it only applies to field UDGs that have a significant HI component (HUDs), which means that this population is distinct, both in terms of its baryonic constituents and its environment, from the UDGs that have been found in clusters, and potentially in its origin as well. In this Section we estimate how number density of this population compares with that of UDGs in groups and clusters, estimate the influence it is expected to have on the calculation of the galaxy stellar mass function (SMF), draw comparisons with current SAMs that produce UDGs, and discuss how it relates to the various recently proposed formation mechanisms for UDGs.

\subsection{Comparison with the number density of cluster and group UDGs}

\begin{figure}
\centering
\includegraphics[width=\columnwidth]{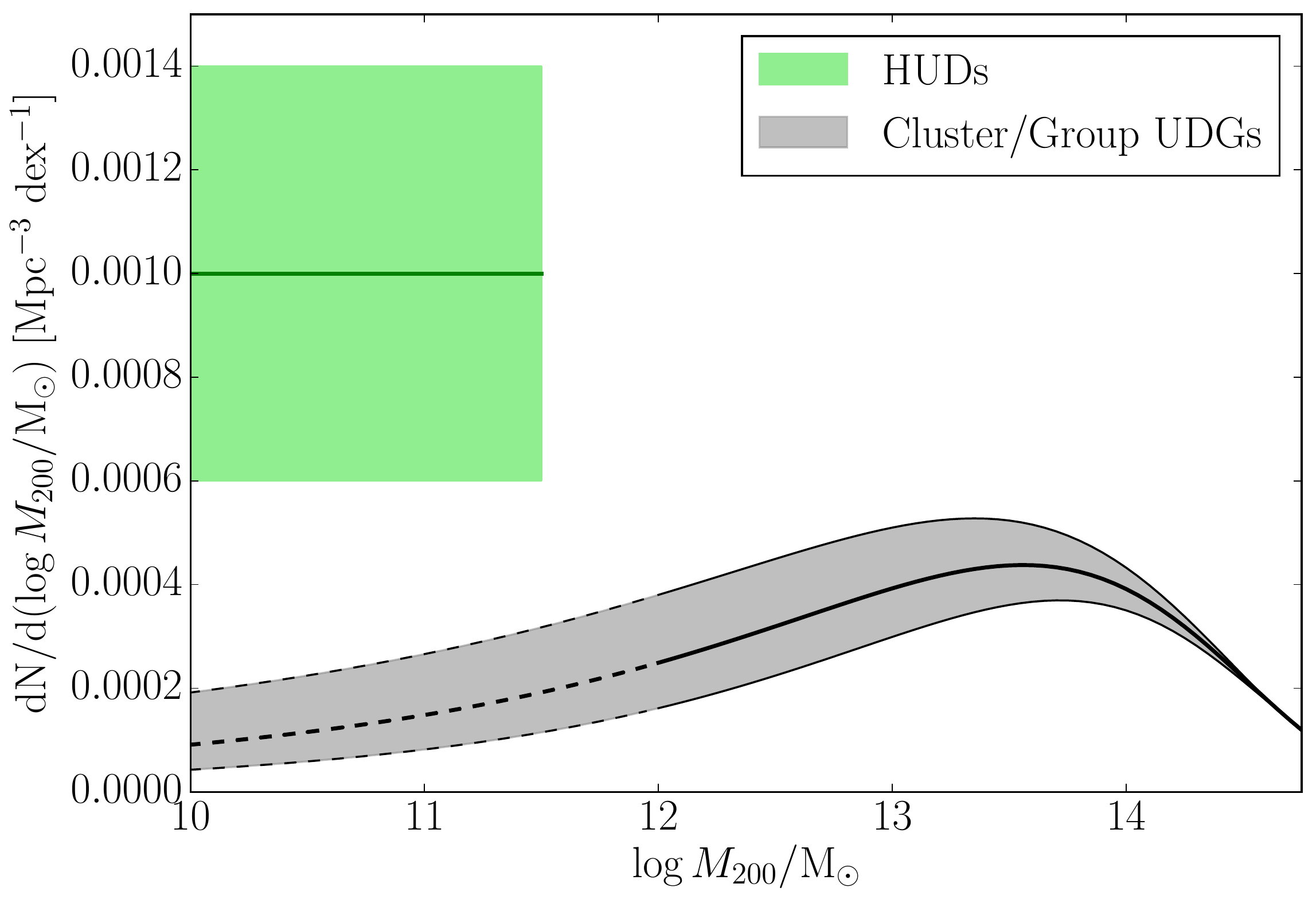}
\includegraphics[width=\columnwidth]{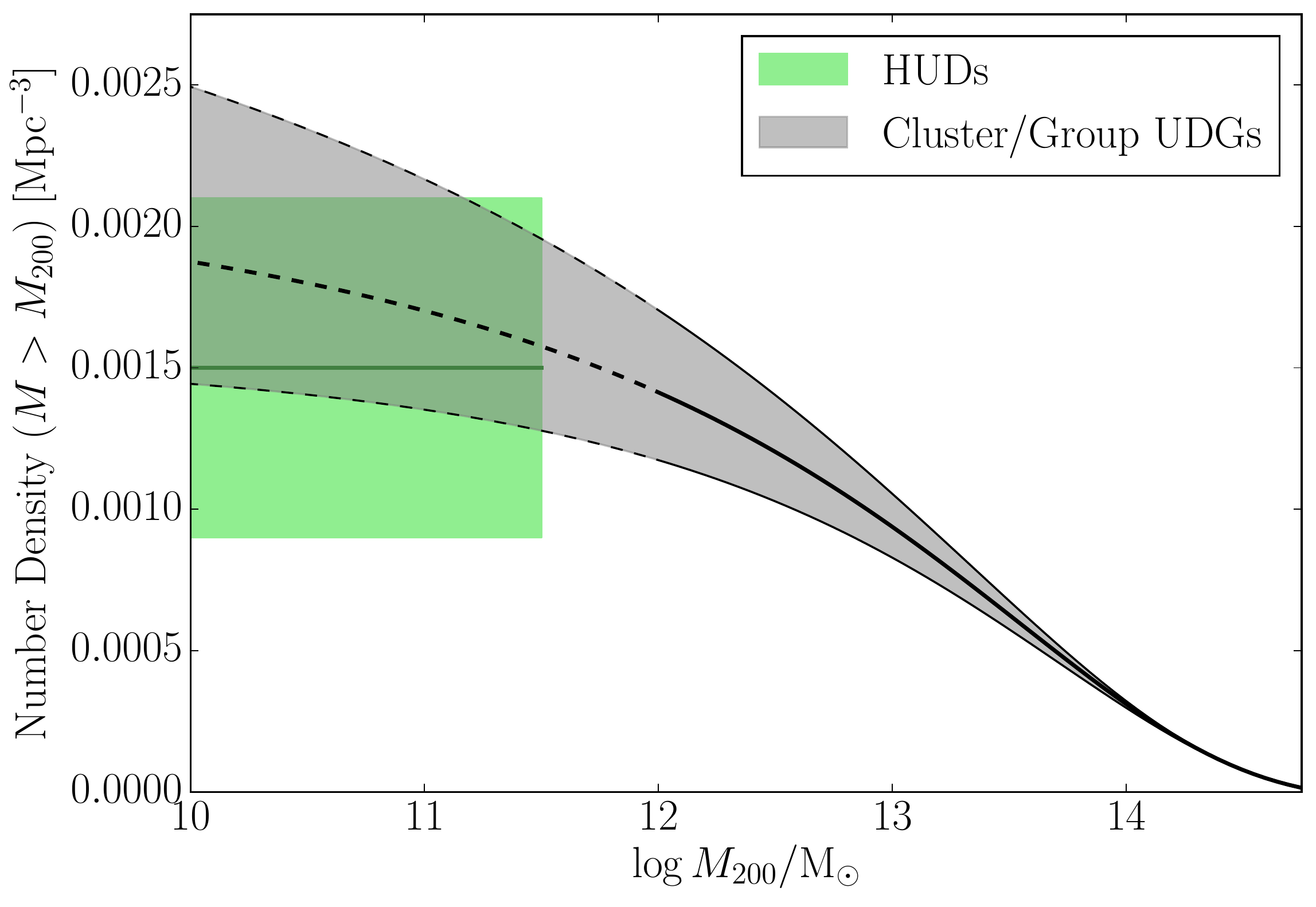}
\caption{
\textit{Top}: The number density of cluster and group UDGs (grey band and black lines) as a function of halo mass. The horizontal green band and line indicates the number density of HUDs assuming a uniform distribution across halo mass $10^{10} < M_{200}/\mathrm{M_{\odot}} < 10^{11.5}$. 
\textit{Bottom}: The cumulative number density of cluster and group UDGs (grey band and black lines) in halos \textit{above} a given mass. The total number density of HUDs (horizontal green band and line) is shown for comparison. 
The central lines and the widths of the bands indicate the central value and 1-$\sigma$ uncertainties respectively. The dashed black lines indicate the region over which the halo mass--UDG number relation has been extrapolated.
}
\label{fig:udg_hud_num_dem}
\end{figure}

\citet{vdBurg+2017} used observations of UDGs in groups and clusters to fit a relation between the mass of the central halo and the number of UDGs that it hosts (their Equation 1). Multiplying this relation by the halo mass function (HMF) gives the differential cosmic number density of UDGs as a function of halo mass. Integrating this function over halo mass thus gives the total cosmic number density of UDGs that are contained in halos in a given mass range.

To perform this calculation we combined the \citet{vdBurg+2017} relation with an analytic approximation \citep{Sheth+2002} for the mass function of distinct halos in the Bolshoi simulation \citep{Klypin+2011}.
\begin{eqnarray}
    n_{\mathrm{UDG}} = \int_{M_{\mathrm{min}}}^{M_{\mathrm{max}}} &1.2 \times (19 \pm 2) \left[ \frac{M}{10^{14}\,\mathrm{M_{\odot}}} \right]^{1.11 \pm 0.07} \nonumber \\
    &\times \, \Omega_{\mathrm{m,0}} \rho_{\mathrm{c,0}} \frac{M f(\sigma)}{\sigma(M)} \frac{\mathrm{d}\sigma}{\mathrm{d}M} \; \mathrm{d}M
\end{eqnarray}
where $M$ is the halo mass, $\Omega_{\mathrm{m,0}} = 0.27$, $\rho_{\mathrm{c,0}}$ is the critical density of the Universe today, and the functions $\sigma(M)$ and $f(\sigma)$ are defined as in Equations B8--B14 of \citet{Klypin+2011}. The factor of 1.2 is because the \citet{vdBurg+2017} relation is based on the $M_{200}$ definition of halo mass, whereas the Bolshoi HMF is for halo virial masses. \citet{Klypin+2011} found that for $M_{200}$ masses the HMF is approximately 20\% higher at a given mass, compared to the HMF for virial masses. We set the maximum halo mass considered to be $10^{15}$ \Msol. When $M_{\mathrm{min}} < 10^{12}$ \Msol \ we extrapolated the \citet{vdBurg+2017} relation (which had no data below this halo mass).

The resulting differential and integrated cosmic number density of UDGs in groups and clusters is shown in Figure \ref{fig:udg_hud_num_dem} along with the cosmic number density of HUDs that we calculated in Section \ref{sec:results}. The bands indicate the 1-$\sigma$ uncertainties on each quantity. In the case of the UDGs, we have assumed that the uncertainties in the gradient and intercept of the \citet{vdBurg+2017} relation are 100\% anti-correlated. As we have very little information on the halo masses of HUDs we have simply assumed that they reside in halos of mass $10^{10} < M_{200}/\mathrm{M_{\odot}} < 10^{11.5}$ and that they are uniformly distributed over this range. This range was chosen because \citet{vDokkum+2017} estimated a range of approximately $10^{10.5} < M_{\mathrm{halo}}/\mathrm{M_{\odot}} < 10^{11.5}$ for the Coma cluster UDGs, while in the simulations of \citet{DICintio+2017} the range was found to be $10^{10} < M_{\mathrm{halo}}/\mathrm{M_{\odot}} < 10^{11}$. Due to the similar optical properties of HUDs and UDGs we make the assumption that they reside in similar mass halos and, in the absence of more information, that they are uniformly spread across this range. It should be noted that if the actual range were to be narrower, or HUDs were not uniformly distributed across it, then the disrecpency shown in the top panel of Figure \ref{fig:udg_hud_num_dem} (discussed below) would increase.

Although the errorbars on both sets of observations are large, these plots indicate that the total abundances of HUDs and (group and cluster) UDGs are similar, but that HUDs are not a straightforward extension of the relation for group and cluster UDGs---they are much more numerous than the relation would predict. This suggests that the relation between halo mass and the number of hosted UDGs has a break in it below the halo mass range probed by \citet{vdBurg+2017}. The relation cannot continue to decrease as we extrapolated or else there would not be a sufficient number of UDGs in lower mass halos to account for all the HUDs.

Cluster and group UDGs are satellite objects, whereas HUDs appear to be field centrals. Therefore, the indication of a break in the relation is not entirely unexpected, and we hypothesise that the break is probably associated with the halo mass at which UDGs transition from being satellite objects, to centrals. Unfortunately, these findings on their own do not illuminate further the formation mechanism(s) of HUDs and UDGs because although they appear to be somewhat separate populations, it is still possible that they have either the same or different formation mechanisms.

\subsection{Contribution to the galaxy stellar mass function}

\begin{figure}
\centering
\includegraphics[width=\columnwidth]{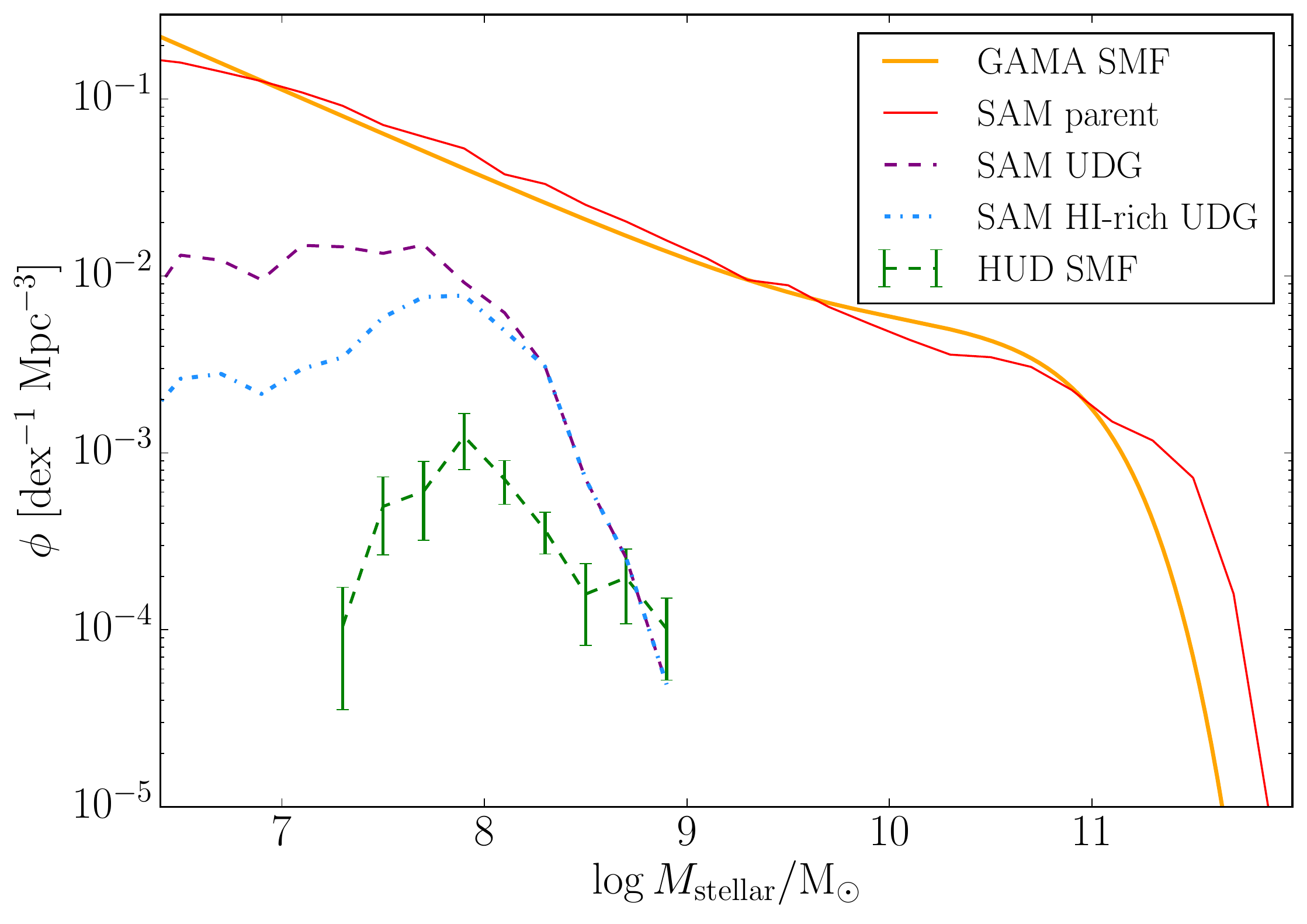}
\caption{
The contribution of HUDs to the galaxy stellar mass function (green dashed line and error bars). Error bars are 1-$\sigma$ and only account for Poisson uncertainties. Bins that contain only one object have been removed. The double Schechter function fit to the GAMA SMF \citep{Wright+2017} is shown for comparison (smooth orange line), as well as the Santa Cruz SAM parent sample SMF (red line), SAM central UDGs SMF (purple dashed line), and the SAM HI-rich central UDGs SMF (blue dash-dot line).
}
\label{fig:hud_smf}
\end{figure}

Without global estimates of the number densities of UDGs it has not been possible to estimate their contribution to the SMF, measurements of which are typically based on spectroscopic samples that do not contain UDGs. One of the most recent measurements of the SMF is from the GAMA (Galaxy And Mass Assembly) survey \citep{Wright+2017}. The authors of that work estimated that the SMF was valid for objects with average surface brightnesses brighter than 24.5 $\mathrm{mag\,arcsec^{-2}}$ in $r$-band. The sample of \citetalias{Leisman+2017} is selected to have average surface brightnesses fainter than 24 $\mathrm{mag\,arcsec^{-2}}$ in $r$-band, placing most of the HUDs sample below the sensitivity limit of the GAMA sample that was used to calculate the SMF. 

We used the revised SDSS photometry of HUDs from \citetalias{Leisman+2017} and the relations of \citet{Roediger+2015}\footnote{Specifically the mass-to-light versus colour relation for $M_r$ given the $g-r$ colour (their table A1) based on the \citet{Bruzual+2003} stellar population (Chabrier initial mass function) models and assuming a value of 4.68 for solar absolute $r$-band magnitude.} to estimate the stellar masses of the galaxies in the sample. Then by making use of the scaled effective volumes for these sources (from the HIMF calculation), we estimated the correction that their inclusion would make to the SMF. 

Caution is required when using these stellar mass estimates as these galaxies may not be well fitted by the \citet{Roediger+2015} relations. They are low stellar mass and their colours and luminosities likely have substantial stochastic contributions because they appear to be dominated by a few bright, blue stars (\citetalias{Leisman+2017}). However, this means that the true stellar masses of HUDs are probably systematically lower than our estimates, which would not change the result shown below.

Figure \ref{fig:hud_smf} shows the contribution that HUDs make to the SMF compared to the GAMA double Schechter function fit. At all stellar masses the correction that would be needed to account for HUDs is around the 1\% level or less. The relative contribution of HUDs to the SMF is even lower than it is to the HIMF because low-mass field galaxies tend to be HI-dominated, and thus the stellar masses of HUDs are even lower than their HI masses, meaning they are translated further to the left in the SMF where there are even higher number densities of normal dwarfs. In addition, the SMF contains all the dwarfs in groups and clusters that are essentially absent from the HIMF because they have had their gas removed. In other words, the inclusion of the HUD population would make a negligible change to the SMF---these objects are simply too rare and their stellar masses are too low.

\subsection{Comparison with the Santa Cruz SAM}

As UDGs are still a relatively recent discovery, one outstanding question about them is simply whether they are formed in existing cosmological simulations and SAMs, and if so, whether they exist in the correct abundance. It is clear from Figure \ref{fig:hud_himf} and Figure \ref{fig:hud_smf} that UDGs are overabundant in the Santa Cruz SAM (with the size re-calibration applied) in terms of both the HIMF and the SMF. In particular, the ratio of the number density of HI-rich UDGs relative to normal galaxies in the SAM reaches a peak of $\sim$20\% at $M_*\approx10^8$ \Msol \ for both the HIMF and the SMF. This ratio is considerably larger than the new observational constraints presented in this paper. It is interesting that while the overall SAM SMF matches the observed SMF quite well \citep[see also][and Yung et al., in preparation]{Somerville+2015}, we do not reproduce the SMF normalization for the special sub-population of HI-rich UDGs (though the shape is similar). The SAM parent HIMF is consistent with that derived in \citet{Popping+2014} for a similar version of the SAM, which also showed an excess at low HI masses. That work also demonstrated that the SAM reproduces the observed $M_{\mathrm{HI}}/M_{*}$ ratio for ``normal'' galaxies down to $M_{*}=10^{7}$ \Msol. This suggests a problem that could be present in other models as well, since ``tuning'' to match the overall galaxy population does not guarantee that the properties of different sub-populations will be reproduced as well \citep[see also][]{Somerville+Dave2015}. 

The comparison of the HIMFs indicates that either the SAM produces UDGs too frequently or that the UDGs that it produces retain too much HI. To the right of the peak in the observed HUD HIMF the SAM would match well if the HI masses of the modelled galaxies were about half what they are, or if such galaxies were created about half as often. The second interpretation, that the SAM's UDGs are too HI-rich, seems unlikely because the majority of the UDGs produced are actually at lower HI masses than the HUDs sample. Also an inspection of the gas fractions of the SAM UDGs revealed that they are broadly consistent with the rest of the dwarf (central) population, rather than being particularly HI-rich. Therefore, the discrepancy appears to be a straightforward overabundance issue.

The downturn in the HUD HIMF at lower masses may indicate that there is a physical process (not reproduced in the SAM) that either prevents the formation of UDGs with very low HI masses, or one that removes their HI, making them essentially invisible to ALFALFA. Alternatively, it is possible that the downturn is an artifact of the $V_{\mathrm{eff}}$ method used to calculate the observed HIMF. When a minimum distance is set (in this case 25 Mpc) the $V_{\mathrm{eff}}$ method can artificially suppress the abundance of the lowest mass galaxies in the sample (see the appendix of Jones et al., in preparation). We used a ratio method to calculate the HUD HIMF (see Section \ref{sec:method}) partly to minimise this effect, but this possibility cannot be entirely ruled out.

One important aspect of our SAM analysis is that we consider central galaxies only, which was done to roughly mimic the isolation criteria applied to the \elc \ sample. This causes the strong downturn in the SAM UDG HIMF at low masses, because satellites have preferentially lower HI masses than centrals and thus their exclusion leads to a deficit of low HI mass objects. While it is tempting to conclude that the lack of a downturn in the $\alpha$.70 HIMF is therefore due to the fact that satellites are not excluded, visual inspection of low-mass ALFALFA sources reveals that most are field objects, not satellites. Furthermore, \citet{Guo+2017} estimated that $\sim$90\% of all ALFALFA sources are centrals, with no dependence on HI mass (although their analysis did not extend to the lowest masses). Thus, a more likely scenario is that this discrepancy is a shortcoming in how the SAM treats HI in low-mass halos (both satellite and central).

A detailed analysis of the physical origin of low surface brightness galaxies and their gas content in the Santa Cruz SAM is deferred to future work. Nevertheless, here we compare the $g-r$ colours of the observed HUDs to the SAM UDGs. Figure \ref{fig:col_comp} shows that the $g-r$ colours of the SAM UDGs match the colours of the observed HUDs quite well. The observed colours form a broader distribution than those from the SAM, but this is likely due to the large uncertainties involved in measuring the colours of these extreme objects using SDSS images (see \citetalias{Leisman+2017}). This agreement is perhaps not surprising given that nearly all of the central UDGs in the SAM are relatively low mass and thus star-forming, like the observed HUDs. HUDs also appear to be a bluer population than some other field UDGs, for example, the UDGs in the lowest density regions covered by \citet{Roman+2017a} have colours $g-r \sim 0.5$. It is also worth noting the that typical $g-i$ colour of a \textit{satellite} UDG in the Santa Cruz SAM is about 0.7, which is similar to the colours of Coma UDGs \citep{vDokkum+2015a}. While this rough colour comparison is promising, a more detailed analysis of SFHs in both the observations and the models is necessary before a meaningful comparison of stellar populations can be made.

\subsection{Formation scenarios}

\begin{figure}
    \centering
    \includegraphics[width=\columnwidth]{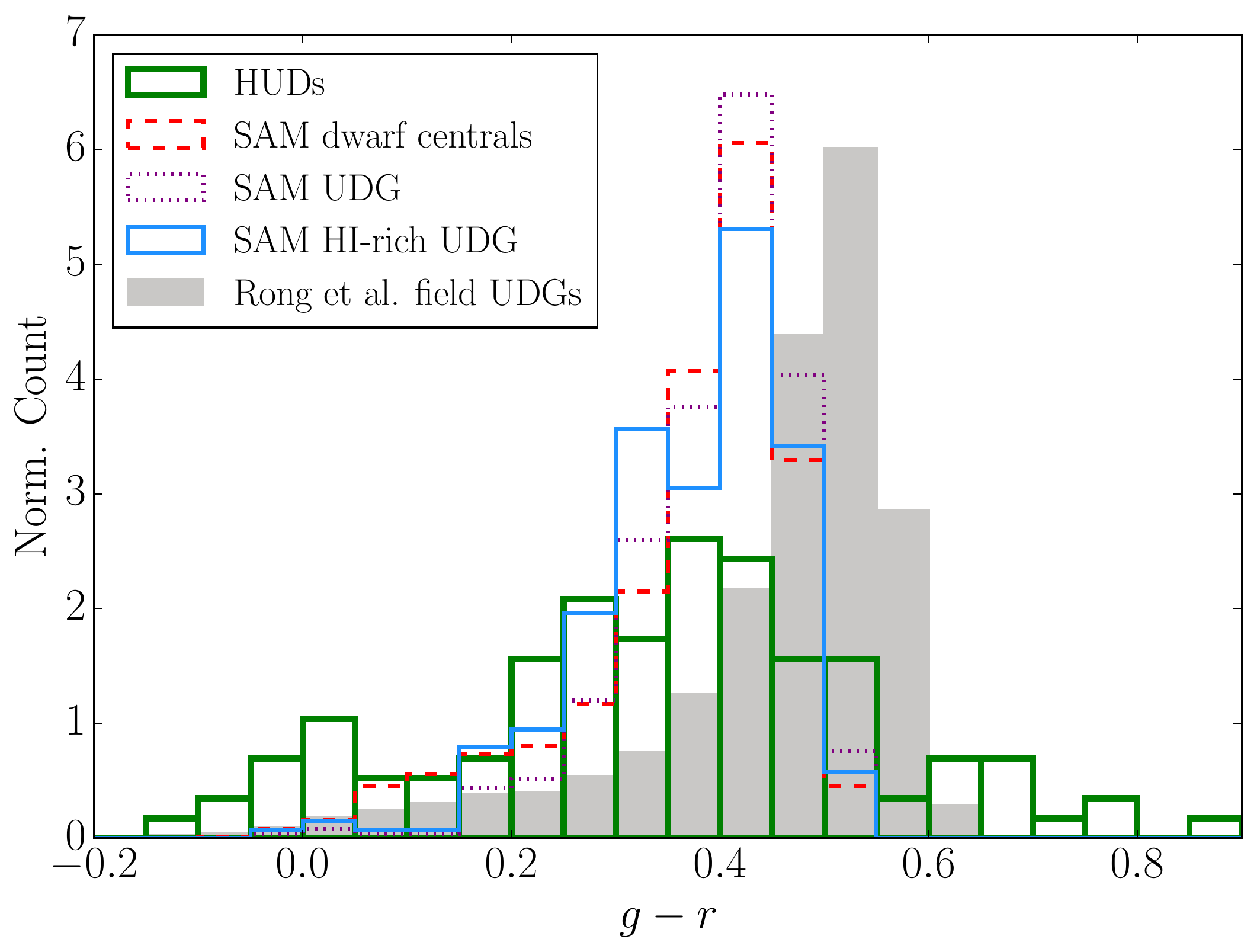}
    \caption{
    The distribution of $g-r$ colours (AB mag) of HUDs (green outline). The dwarf ($M_{*} < 10^{9}$ \Msol) centrals from parent population of the Santa Cruz SAM are shown with the red dashed outline, the central UDGs with the dotted purple outline, and the HI-rich central UDGs with the solid blue outline. The field UDGs discussed in \citet{Rong+2017} are also shown as the grey filled bars. 
    }
    \label{fig:col_comp}
\end{figure}

\citet{vDokkum+2015a} initially proposed that UDGs could be failed $L^{*}$ galaxies, and this idea was developed further by \citet{Yozin+2015}. More recently proposed formation scenarios have preferred the interpretation that UDGs reside in dwarf-sized halos, with the mechanisms driving the galaxies to become ultra-diffuse falling into two categories: those where high halo spin parameters prevent objects becoming more condensed \citep[e.g.][]{Amorisco+Loeb2016,Rong+2017}, and those where stellar feedback disperses matter, creating a diffuse galaxy \citep[e.g.][]{DICintio+2017,Chan+2017}. In both scenarios gas-bearing UDGs are expected to exist in the field.

\citet{DICintio+2017} used the NIHAO \citep[Numerical Investigation of a Hundred Astrophysical Objects,][]{Wang+2015} zoom-in simulations to argue that UDGs can be formed when repeated episodes of star formation cause outflows from low-mass galaxies that drive matter (both baryonic and dark) to larger radii. This is in contrast to the typical dwarf galaxies in their simulations that have a single major star formation event early in their lifetime that permanently removes the majority of their gas, leaving them relatively gas-poor (but still star forming) and more centrally concentrated. Within this model UDGs are expected to exist in all environments, have younger stellar populations than normal dwarfs, and, crucially, should contain neutral gas (when in the field). Their simulated UDGs have $7 < \log M_{\mathrm{HI}}/\mathrm{M_{\odot}} < 9$ with a mean of 8.4. This is qualitatively similar to our population of HUDs, although we cannot see the lowest mass sources owing to the minimum distance cut that we apply. On the other hand, the ratios of UDGs to normal dwarfs appear to be very different, since in the simulations the UDGs are more common than normal dwarfs while they are a very small fraction among the observed galaxies.

\citet{Chan+2017} also identified a stellar feedback driven formation mechanism in their FIRE-2 (Feedback In Realistic Environments) zoom-in simulations. Their objects were mainly hosted by dwarf-sized halos, but some would have grown to $L^{*}$-sized halos had they not been quenched upon entering a cluster environment. They also did not have high spin parameters. In addition, \citet{Chan+2017} found that low stellar mass ($M_{*} < 10^{8}$ \Msol) galaxies remained ultra-diffuse even if they were never quenched, again indicating that a blue UDG population would be expected in the field for a larger simulation.

Unfortunately, a detailed comparison of the number density of field UDGs arising in hydrodynamical simulations is not possible at this stage because no (current) hydrodynamical simulation produces UDGs within a cosmological volume.
In addition, in a stellar feedback driven formation scenario UDGs are not expected to occupy especially high spin parameter halos, which may be in tension with the results of \citetalias{Leisman+2017} that suggest HUDs may be hosted by high spin parameter halos. However, it should also be noted that the recent work of \citet{Spekkens+2017}, which performed HI observations of five blue UDGs identified by \citet{Roman+2017}, indicates the opposite result, that HI-bearing UDGs may not be in high spin parameter halos. Caution is advised when drawing conclusions related to halo spin parameters due to the many assumptions that are involved in their calculation from observables, particularly because some may not apply to low-mass galaxies \citep{Hernandez+2007,El-Badry+2017}.

\citet{Amorisco+Loeb2016} proposed that UDGs may simply be the natural extension of the dwarf population into the high valued tail of the halo spin parameter distribution, with the high halo spins causing the galaxies to be more diffuse. \citet{Rong+2017} found a similar scenario in SAM of \citet{Guo+2011}, based on the Millennium-II and Phoenix simulations, although they pointed out that in addition to residing in high spin parameter halos, their model UDGs also appear to have formed more recently than other dwarfs, which might explain their relatively low abundance in the centres of clusters and groups, and their lack of tidal disruption. As this formation mechanism relies on the spin parameters of the parent halos, not environment, in this scenario UDGs are also expected to exist in the field as well as in clusters.

The results of this SAM indicate that the number density of field UDGs should be of the order of $10^{-2}$ Mpc$^{-3}$ (Yu Rong, private communication). This is approximately an order of magnitude above what we have estimated for HUDs, and indicates that the \citet{Guo+2011} SAM has a similar over-production issue as the Santa Cruz SAM (discussed above). \citet{Rong+2017} also predicted that about 14\% of all field dwarfs should be UDGs. This appears inconsistent with our findings that at most $\sim$6\% of HI-bearing dwarfs can be classified as ultra-diffuse. However, this apparent tension might be resolved if a substantial fraction of the real field UDG population is devoid of neutral gas or the HUD sample is too restrictive in its optical selection criteria (see Section \ref{sec:caveats}), as either of these would mean that the HUDs sample is missing field UDGs. Finally, a comparison of the colour distributions (Figure \ref{fig:col_comp}) shows that although the \citet{Rong+2017} field UDGs are bluer than quenched dwarfs, HUDs are still generally about 0.2 magnitudes bluer. It should be noted that the colours of HUDs calculated by \citetalias{Leisman+2017} are based on the difference of the measured central surface brightnesses in the two bands, as this was deemed to be more reliable than the magnitude. Combining in quadrature the estimated uncertainties in $\mu_{g,0}$ and $\mu_{r,0}$ (from \citetalias{Leisman+2017}) gives the typical uncertainty in the $g-r$ colour as approximately 0.15, which explains the broad shape of the HUD colour distribution. The different colour distributions of UDGs in the Santa Cruz and \citet{Guo+2011} SAMs is likely because colours are quite sensitive to both dust and the level of ``burstyness'' in the star formation histories, which can be quite different in the two SAMs as they use different treatments of star formation.

The clear tendency of HUDs to have narrow velocity widths, relative to other HI-rich dwarfs, might be an indication of their formation mechanism. If we make the assumption that the HI distribution is at least as extended as the distribution of stars (which is almost always true for other types of HI-rich galaxies, and for the few HUDs for which there are interferometric 21 cm observations, \citetalias{Leisman+2017}), then the typical argument used to explain narrow velocity widths in other dwarf galaxies, that only the centre of the DM distribution is being traced by the gas, is not applicable. Two possible physical explanations could be that either the DM halos of HUDs are lower mass than those of other dwarfs (of the same HI mass), and have correspondingly low maximum circular velocities ($V_{\mathrm{max}}$), or they are less centrally concentrated than normal, leading to gradually rising rotation curves. While the interpretation that the DM halo might be less centrally concentrated seems to favour the \citet{DICintio+2017} feedback driven formation model, lower central concentration (at a given $V_{\mathrm{max}}$) will also mean a higher spin parameter, which could favour the formation scenario proposed by \citet{Amorisco+Loeb2016}. An alternative explanation is that the prevalence of low rotation velocities in HUDs might simply be due to a selection bias in favour of either slowly rotating sources or low inclination sources. While this possibility cannot be entirely ruled out, we think it is unlikely (as is discussed in the following sub-section).

Ideally one would be able to use interferometric observations to trace the rotation curves of HUDs in order to reveal the density distributions and masses of their DM halos. However, HUDs (as is apparent from Figure \ref{fig:hud_wf}) generally have extremely narrow velocity widths such that ordered rotation is difficult to untangle from random turbulent motions. Another observational difference predicted by the proposed formation scenarios is the morphology of UDGs: whether they are more disc-like or spheroidal. The \citet{DICintio+2017} outflow mechanism creates more spheroidal UDGs, whereas the high spin parameter model \citep{Amorisco+Loeb2016,Rong+2017} produces disc-like UDGs. Unfortunately, as the light distribution of HUDs is very patchy and typically dominated by recent star formation, morphologies and inclinations are extremely challenging to determine, and at present we cannot distinguish the models in terms of morphology.

\subsection{Caveats when comparing to simulations and models}
\label{sec:caveats}

Although in the above discussion we have made direct comparisons with values from models, there are some important caveats regarding such a comparison.

The first is regarding the main result of this paper, the cosmic number density of HUDs. The criteria we used to define what is a HUD were intentionally the same as what has been used to define a UDG, specifically the definition used by \citet{vdBurg+2016}. However, while the cluster UDGs are typically red and appear devoid of gas and star formation, HUDs are bluer, clearly have ongoing star formation, and contain HI by definition. Thus, for UDGs a threshold surface brightness in $r$-band should convert straightforwardly into a threshold in stellar mass surface density (the physical property of interest), whereas for HUDs this conversion is less straightforward, both because the stellar population is not red and dead, and because it is not smoothly distributed. As both of these effects will act to increase the surface brightness of the objects we suggest that although this population is directly analogous to the cluster UDGs (in terms of their selection criteria), the truly analogous field population would likely include many more dwarfs that are considerably brighter. To correctly identify this population would require accurate stellar masses for all the dwarfs in ALFALFA. This presents a significant challenge as the automated photometry for many of these sources is not adequate and standard models of stellar populations are not representative of these low-mass, gas-rich objects \citep[e.g.][]{Huang+2012}.

Another caveat concerns the colours of HUDs. Owing to the challenging nature of the sources, \citetalias{Leisman+2017} simply forced an exponential fit within circular apertures in order to calculate magnitudes. This leads to uncertainties in the colours that are difficult to quantify and potentially large. Having said this, a visual inspection of HUDs confirms that they are clearly blue, while the field UDGs of \citet{Rong+2017} have similar colours to UDGs found in clusters.
However, the fact that the colours of HUDs are not always reproduced by current models of formation scenarios is perhaps not so much a disagreement with the mechanisms causing sources to become ultra-diffuse, as it is with simulations and SAMs in general. As discussed in \citet{Somerville+Dave2015} it is an almost universal problem with existing simulations and SAMs that the low-mass galaxies that they produce tend to form the majority of their stars much earlier than real dwarf galaxies appear to. Thus, the fact that observed HUDs are bluer than the modelled sources in \citet{Rong+2017} is likely not a problem that is unique to UDGs. It is, however, a cautionary note for any formation mechanism that depends on the specifics of the star formation history.

It is also worth noting that the size--mass relation used in the Santa Cruz SAM is calibrated on that of the GAMA survey \citep{Lange+2015} which does not extend to the low stellar masses of HUDs. Therefore, the relation is necessarily extrapolated when modelling the properties of the most extreme objects produced in the SAM. As the observational relation between magnitude or mass and the radii of UDGs is steep \citep[e.g.][]{Koda+2015,vdBurg+2016} a relatively small change in the size--mass relation used in the SAM could have a large effect on the inferred number density. This was discounted as a source of disagreement between the observed number density of HUDs and central UDGs in the SAM because Figure 6 of \citet{Lange+2015} suggests that there may be an upturn in the GAMA relation at low masses, which would lead the SAM to under-produce, not over-produce, UDGs. However, as this upturn occurs below the completeness limit of GAMA a change in the oppose sense cannot be completely ruled out. Such a change in the relation could be the source of some of the disagreement between the models and the observations of HUDs. This will be investigated further in another paper.

The steep increase in the HUD velocity width function towards narrow widths might be taken as evidence of a bias for selecting almost face-on systems as HUDs. In terms of the HI selection, narrow velocity width (i.e. almost face-on) sources are more easily detectable because their 21 cm flux is spread over a narrower frequency range, leading to a higher peak flux density relative to the noise. However, this is a well understood effect that is incorporated in the shape of the ALFALFA detection limit \citep{Giovanelli+2005,Haynes+2011} and the $V_{\mathrm{eff}}$ method makes an implicit correction for it. We therefore find it unlikely that this is a source of bias. 

In the case of the optical selection criteria, a face-on source would naively be expected to have lower surface brightness, making a given object more likely to be classified as ultra-diffuse. However, as the SDSS images are not especially deep and the light distributions of HUDs are patchy, in some cases it may be possible that much of the area of a face-on source is not detectable in SDSS. This might lead to a large underestimation of radii, based only on the brightest knots of star formation, which would mean they are omitted from the final sample. As inclinations of these sources are extremely difficult to measure, the scale of these uncertainties are difficult to estimate, and they will only be solved with improved imaging.

Owing to issues such as the ones discussed in the previous paragraph, the exact HUDs sample size is uncertain. There are fixed criteria that determine if a source is a UDG, but the SDSS photometry which was used to assess if the criteria were met, had large uncertainties. In total \citetalias{Leisman+2017} estimated that the 1-$\sigma$ variation in the HUDs sample size was about 25\%, and this uncertainty is not modelled in our estimates of the HUD HIMF, WF, and SMF. While 25\% is a substantial uncertainty the current discrepancy between observations and models is considerably larger. Therefore, we note this issue but do not attempt to account for it because it would not alter our qualitative results. Furthermore, the Poisson uncertainties in the number density and HI mass density of HUDs are at or above this level of uncertainty already.

A final note of caution is that while HUDs are certainly an interesting population, there remains the possibility that the discrepancy in number density we are finding between HUDs and field UDGs in SAMs is not only because of shortcomings in how HI is modelled for low-mass objects, but also because HUDs might only be the tip of the iceberg of the field UDG population. Indeed, the preliminary sample of field UDGs from the Hyper Suprime-Cam survey \citep{Greco+2017} was matched to ALFALFA, finding matches in only $\sim$1\% of cases (however, without redshifts it is difficult to know whether ALFALFA is expected to detect these sources or not). The potential existence of a population of HI-poor and quiescent dwarf galaxies, essentially invisible to ALFALFA and current wide-field optical surveys, was put forward by \citet{Giovanelli+2015}---the so called Too Shy To Shine population. If such a population exists, it is possible that it might have avoided detection by optical surveys thus far if the dwarfs were ultra-diffuse. Therefore, the only way to identify (or rule out) such a population is to carry out deep optical surveys that blindly cover the field. Such a survey is underway with Hyper Suprime-Cam \citep{Greco+2017}, although the necessity of accurate redshifts for a robust accounting of number density will remain a challenge for the immediate future.

\section{Conclusions}
\label{sec:conclusions}

We have used the sample of HI-bearing ultra-diffuse sources (HUDs) identified in ALFALFA (\citetalias{Leisman+2017}) to calculate the first estimate of the cosmic abundance of UDGs in the field, which we find to be $(1.5 \pm 0.6) \times 10^{-3}$ Mpc$^{-3}$. They form a small fraction of the overall HI population, peaking at a fractional abundance of about 6\% of galaxies with HI masses of $\sim$10$^{9}$ \Msol. Their velocity widths are found to be much narrower on average than the parent HI population, with the distribution rising steeply from a maximum value of $\sim$100 \kms \ to a minimum value of $\sim$20 \kms. We estimate that these sources, which are too low surface brightness to be included in most spectroscopic samples, represent less than a 1\% correction to the galaxy stellar mass function.

Using the halo mass--UDG number relation of \citet{vdBurg+2017} and the HMF \citep[from the Bolshoi simulation,][]{Klypin+2011} we estimated the total cosmic abundance of group and cluster UDGs in halos above $M_{200} = 10^{12}$ \Msol \ to be $(1.5 \pm 0.3) \times10^{-3}$ Mpc$^{-3}$. This indicates that the abundance of field UDGs is likely at least comparable to the abundance of those in clusters and groups, and we interpret this finding as an indication that the halo mass--UDG number relation cannot continue unbroken at halo masses below $10^{12}$ \Msol, otherwise the field population would be considerably sparser.

The population of UDGs produced by the Santa Cruz SAM was compared with the HUDs from ALFALFA. While the modelled HI-rich UDGs have a similar colour distribution to HUDs, likely indicating similar stellar populations, they are produced in much greater abundance than the observed abundance of HUDs. We also compared the properties of HUDs to those predicted based on current proposed formation scenarios. Both the mechanisms of UDGs forming in high spin parameter halos or due to repeated episodes of stellar feedback predict there to be HI-rich UDGs in the field, which the \citetalias{Leisman+2017} sample confirms. A more detailed comparison was only possible for \citet{Rong+2017}, which discusses the UDGs created by the \citet{Guo+2011} SAM. Their field UDGs also suffer from an over-production problem, occurring about 10 times more frequently that the HUDs found in ALFALFA. They also make up a larger fraction of the modelled dwarf population, and have colours that are too red (a common problem for dwarf galaxies in SAMs and simulations). 

These findings idicate that SAMs currently produce field UDGs too readily. Alternatively, some of the tensions might be resolved if the the HUD population is a highly incomplete census of the field UDG population, that is, if most field UDGs are not HI-rich or many UDGs have been excluded from the HUD sample due to bright star formation knots that violate the optical surface brightness criteria. However, the former would require that a significant fraction of the field dwarf population has been missed by all optical surveys thus far.

At present SAMs appear unable to accurately recreate the UDG population, with modelled field UDGs either appearing too frequently, with too much neutral gas, or with colours that are too red. However, this is unsurprising given the relatively recent discovery of this population. HUDs represent a complementary sample to the UDGs found in clusters, with different morphologies, colours, baryonic content, and abundance. Together these properties will provide constraints for models and simulations that will allow improvements in the modelling of UDGs to be assessed over the entire population, both in clusters and in the field.

\begin{acknowledgements}
We thank Yu Rong for kindly providing the colours and number density of their modelled field UDG population.
MGJ acknowledges support from the grant AYA2015-65973-C3-1-R (MINECO/FEDER, UE).
EP has been supported by a NOVA postdoctoral fellowship at the Kapteyn Astronomical Institute.
AJR was supported by National Science Foundation grants AST-1616710 and PHY11-25915, and as a Research Corporation for Science Advancement Cottrell Scholar.
RSS is grateful for the generous support of the Downsbrough family, and acknowledges support from the Simons Foundation in the form of a Simons Investigatorship.
EAKA is supported by the Women in Science Excel (WISE) programme, which is financed by the Netherlands Organisation for Scientific Research (NWO).
\end{acknowledgements}

\bibliographystyle{aa}
\bibliography{refs}

\end{document}